# Highly photo-stable, kHz-repetition-rate, diode pumped circulation-free liquid dye laser with thermal lens management

A. Hamja,[a)] R. Florentin, S. Chénais, S. Forget

Laboratoire de Physique des Lasers, Université Sorbonne Paris Nord, CNRS, UMR 7538, F-93430 Villetaneuse, France

[a)] Author to whom correspondence should be addressed: mdamir.hamja@sorbonne-paris-nord.fr

**ABSTRACT**

Liquid dye lasers have long been considered as ideal tunable laser sources in the visible range, but are bulky, expensive and require a complex system for dye circulation. Here we present a system that relies on a low-cost blue laser diode as the pump source and a sealed dye cell with no flowing circuitry, resulting in a device that combines the convenience and size of a solid-state device with the stability of a liquid organic laser. A very high photo-stability is obtained (up to $1.2\times10^9$ pulses, or 12 days at 1 kHz), which is 5 orders of magnitude higher than a solid-state dye laser operated in similar conditions. The number of pulses obtainable at low repetition rates is found to be limited by molecular self-diffusion, and hence related to the total cuvette volume. In contrast, the repetition rate is limited to a few kHz which suggests that thermal effects play a bigger role than triplet population effects. Thermal effects participate in the suppression of lasing through the buildup of a strong negative thermal lens: correcting the non-aberrant part of this thermal lens by resonator design enables the repetition rate to be pushed up to 14 kHz, with possible further optimization. This work shows a route for building off-the-shelf, compact, low-cost, convenient tunable pulsed lasers in the visible range that have superior stability over organic solid-state lasers.

Recent advent of high-power blue and violet laser diodes[1] has contributed to reconsider many applications that formerly required expensive solid-state lasers in the visible: for instance, Titanium-sapphire lasers' cost can be dropped by one order of magnitude when pump lasers are replaced by laser diodes.[2] Because they exhibit large absorption bands in the blue-green region of the spectrum, dye lasers are also well adapted for GaN diode-pumping and might follow the same route.[3–8] However, while liquid dye lasers have been the first tunable lasers, useful for *e.g.* spectroscopy,[9] medicine,[10] or sensing,[11] they are nowadays more confidential, mostly because of a painful and complex handling of dye solution circuitry. Indeed, in those lasers, the gain medium has to be continuously replenished by an active flow: this avoids accumulation of triplet states, relaxes thermal problems and evacuates photo bleached molecules to enable stable lasing. The complexity of dye circuitry is a bottleneck for many applications, although it can be somewhat minimized by the use of optofluidic devices.[12,13] In order to overcome this difficulty, two solutions can be implemented: solid-state gain media or circulation-free liquid capsules. While solid-state dye lasers are considered as very promising (specifically after the advent of organic semiconductors which triggered the hope to build an electrically-pumped organic laser diode[14–16]), they suffer from important drawbacks. The main one is a low photo-stability and an inability to operate at high repetition rates : despite numerous efforts to improve this issue, such as enhanced chemical design,[17] addition of triplet quenchers,[18] encapsulation[19] or mechanical motion[20] of the gain medium, a typical upper limit for stability is met around $10^6$ pulses[21–24] for pump durations in the nanosecond range, and typical repetition rates for static systems that stand around a few Hz.[25] A notable but unique exception is found in a low power quasi-Continuous-Wave DFB laser demonstrated using BSBCz dye where lasing life-time was, in the order of ~$10^9$ pulses at 80 MHz, limited by thermal degradation.[26]

Diode-pumped circulation free liquid dye lasers[3–8] are consequently an option to be considered, as a valuable way to gather ease of operation with improved stability and higher repetition rate. Such devices represent a new opportunity to enter the realm of efficient, low-cost, compact laser systems, well-suited for real-world applications. However, reports about their photo-stability and maximum repetition rate are scarce in the literature.

In this study, we report on the stability performance of an organic laser in which the gain medium is a dye solution dispensed in an off-the-shelf spectrophotometric cuvette (0.5 ml), inserted into a Plano-concave resonator based on commercially available components, and longitudinally pumped by a couple of blue laser diodes (445 nm). This design makes this laser very simple, with no technological step for device fabrication, and no circulation tubing. Simplicity of the setup makes the laser attractive for any research application that requires low-power pulsed light at an "unconventional" wavelength not accessible with standard lasers. The easy replacement of the gain medium also makes it convenient for screening new organic dyes to evaluate their lasing potential directly from a solution.[27] It is also valuable for educational purposes as the laser structure can be built from scratch with the nice advantage of having both pump and laser wavelengths in the visible range.

The laser structure is detailed in Fig. 1. The pump head combines two polarization-coupled laser diodes with 5W nominal CW power at 445 nm and pulsed through an electronic driver (PCO-7120 from Directed Energy, Inc.) to provide adjustable pulse duration and repetition rate. In order to suppress ellipticity and astigmatism, the diode beams were coupled to a multimode fiber ($r_{core} = 100$ µ$m$ core radius, numerical aperture NA=0.22) and then focused onto the cuvette to create 60 µ$m$ spot radius. The gain medium was a spectrophotometric cuvette 11.8 mm long (easily adjustable) and closed by a concave output coupler (R ~ 99.9% at 600-680 nm) with a 50 mm radius of curvature. The fundamental cavity mode being ~ 80 µm in radius, a good matching is realized with the pump mode along the cuvette light-path. Lasing is obtained with a diffraction limited TEM$_{00}$ beam, and a threshold of 47 kW/cm$^2$ (**Fig. S1(a,b) in supplementary material**). The total footprint of the actual setup is 29 cm by 18.5 cm and could easily be made much smaller. The cavity is designed here to minimize the threshold, resulting in low conversion efficiency (<1%, see **Fig. S1(a)**). However, we had shown in a previous study that a similar setup can yield a conversion efficiency of ~ 18% when the output coupler reflectivity is reduced to R~98%.[3]

The device lifetime was measured using 0.03 mL of DCM solution, 2.4 times above threshold at a pump power density of 122 kW/cm² (pump pulse duration 50-ns). The photo-stability curves are shown on Fig. 2. The liquid dye device stopped lasing after $6\times10^8$ pulses (half-life time is around $3\times10^8$ pulses) with pulse-to-pulse stability around 5%, corresponding to more than 166 hours (approx.

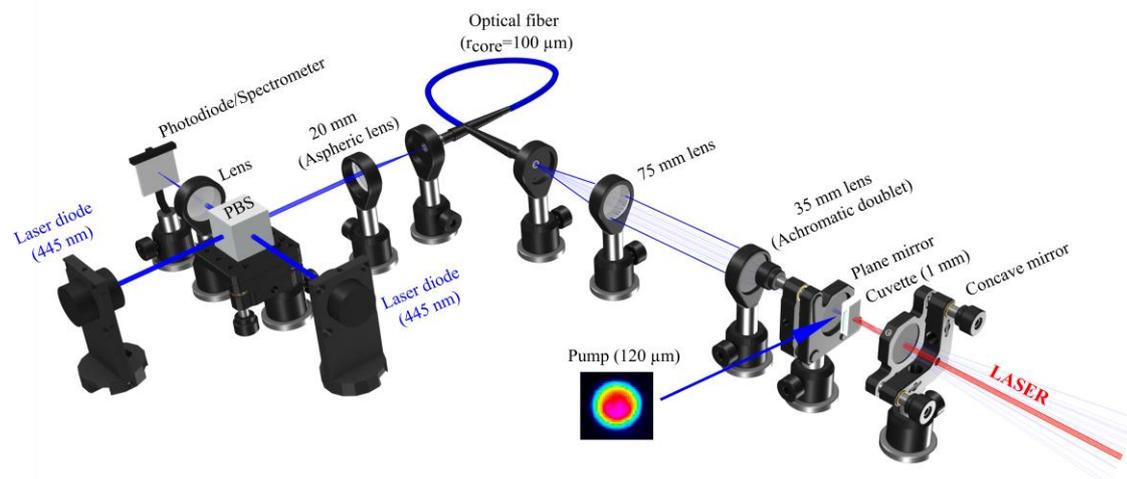

Fig. 1: Schematic diagram of experimental setup (PBS: Polarization beam splitter).

(Hellma analytics) filled with a solution of 4-Dicyanomethylene-2-methyl-6-p dimethylaminostyryl-4H-pyran (DCM) in propanol-2 ($1.76\times10^{-4}$ M, 84% absorption at 445 nm). The cuvette had a 1-mm light path length, in order to be on one hand not too large compared to the Rayleigh distance of the pump circular beam after the fiber ($Z_R = w^2/(r_{core}.NA) = 0.16$ mm) and on the other hand not too slim to enable good absorption at dye concentrations well below the solubility limit. The cuvette is set as close as possible from the plane mirror (reflectivity R ~ 99.9% at 600-680 nm, transmission T > 80% at 445 nm); the cavity is a week) of non-stop lasing at 1 kHz. We then filled the cuvette with more dye solution (approx. 0.3 ml), which increased the laser lifetime up to $1.3\times10^9$ pulses (**Fig. S2**) (half-life time around $7.5 \times 10^8$ pulses) corresponding to more than 360 hours of non-stop lasing at 1 kHz. In this case, we noticed that the maximum laser output was reached only after *ca.* $10^8$ pulses, which is correlated to an increase of absorption of the sample and a small reduction of the solution volume with time, which we attribute to some evaporation of propanol through the cuvette sealing lid. In either case, we note that loss of lasing after complete photo-degradation is irreversible and cannot be recovered at any pump spot location on the cuvette. Together with the observed higher stability for the largest cuvette, this result shows clearly that molecular diffusion is active to replenish the photo-excited volume with fresh molecules continuously between pulses.

The photo-degradation of the sample was characterized by comparing the absorption of the dye solution before and after illumination (**Fig. S3**). In addition of the $S_0 \rightarrow S_1$ absorption peak of DCM in propanol-2 at 470 nm, a second absorption peak around 348 nm emerged in the degraded sample. This characteristic higher energy absorption band was reported earlier,[28,29] and was attributed either directly to the DCM *cis*-isomer or to a degradation photoproduct preferentially created from the *cis*-isomer, whose formation from the most stable *trans*-isomer is a distinguishing photo-physical process occurring in stilbene dyes.[29,30]

As a comparison, we used the same setup to pump a solid state sample: to this end, a 17-µm-thick layer of PMMA doped with DCM was directly coated on the plane mirror,[31,32] with ~ 83% absorption at the pump wavelength. The solid-state laser degraded much faster (half-life of 1850 pulses only) when pumped at same peak power density (122 kW/cm$^2$). Let's note that comparison with solid-state is not straightforward, as the dye concentration was much higher in the solid than in the liquid form (1.57×10$^{-1}$ M *vs.* 1.76×10$^{-4}$ M) in order to achieve high enough absorption along a shorter path length. These results however confirm a well-established order of magnitude for solid-state laser stability. Another comparison point can be taken from the work of Hu *et al.*[33] who reported DCM dye laser with a 3-mm thick solid-state film using ORMOSIL gel, where the dye concentration was 4×10$^{-4}$ M, is very much comparable to the concentration we used. They showed a 10% decrease of lasing output energy after only 2×10$^4$ pulses, a drop in intensity which happens in our case after 5×10$^7$ pulses.

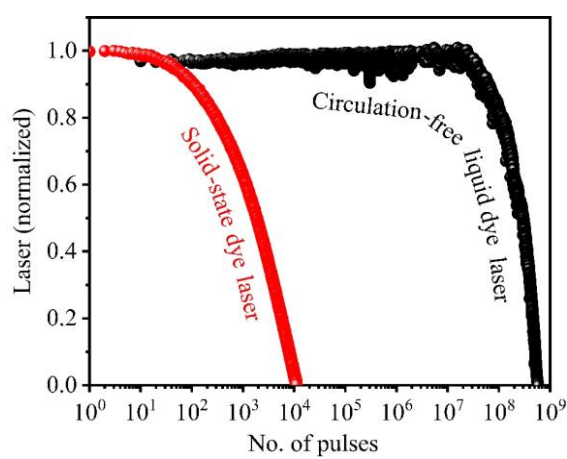

**Fig. 2.** a) Photo-stability of circulation-free liquid dye laser (using 0.03 mL DCM solution in 1-mm light path cuvette) and solid-state dye laser (using 17-µm DCM film) at ~ 122 kW/cm$^2$ pump (50-ns, 1 kHz) power density. The absorption of both liquid and solid-state samples was around 83% at 445 nm.

In addition to this extended lifetime of several orders of magnitude, our circulation free liquid dye laser allows operation at much higher repetition rate: while solid-state lasers are often operated at repetition rates of a few Hz to enable sustained lasing for at least a few seconds with a manageable thermal load, the liquid laser can here operate at 1 kHz. Upon increasing the pump repetition rates, we however observed that the laser efficiency rapidly dropped beyond 1 kHz, with no lasing beyond 2.4 kHz under 30-ns pumping. This demonstrates a cumulative effect (each pulse being influenced by a perturbation brought to the medium by the previous pulses) that has a typical *ms* timescale. We exclude triplet states from being directly responsible of this limit: indeed, even if triplet lifetime can be in the *ms* range or longer in the solid-state, they are typically much shorter in liquids, especially when solutions are not deaerated to eliminate molecular oxygen which acts as an effective triplet quencher. Meyer *et al.* measured a triplet lifetime of ~ 31±5 µs[29] of DCM in a deaerated DMSO solution. We consider here this value as an upper limit, as our solution was not oxygen-free, indicating that the cut-off frequency is rather to be expected in the 0.1 MHz range for triplet contributions. Therefore, thermal effects are the most likely limiting factor, in particular thermal lens diffraction losses[34,35] that increase with pump repetition rate[3] and tend to push the cavity outside of its stability zone. The most obvious effect that can be expected is indeed a pump-induced heating that creates an index profile which distorts the phase front of an incident plane wave, the quadratic component of this phase shift being ascribable to a thin lens with a focal length $f_{th}$.

A simple pump-probe experiment was set up to obtain an order of magnitude for the steady-state thermal lens focal length at different pump repetition rates. The same experimental arrangement (fig.1) was used but the cavity mirrors were removed. Details of the pump-probe experiment is given in **Fig. S4**. Briefly, a He-Ne laser probe (632.8 nm) was made to travel through the pump spot location and changes in probe beam radii with the increase in pump (30-ns, 77.4 kW/cm$^2$:1.6 times above laser threshold) repetition rates were monitored at a fixed distance (7.7 cm) from the cuvette.

Results appearing in Fig. 3 show a clear increase in probe beam radii as a function of pump repetition rates, which can be ascribed to the formation of a diverging thermal lens (as the *dn/dT* coefficient of

propanol-2 is negative).[36] If we assume that the observed probe beam enlargement only results from a perfect non-aberrant thermal thin lens, one can infer the thermal lens dioptric power by using the ABCD ray-matrix method to simulate the experimental results. When the repetition rate was increased from 0.3 kHz to 5 kHz, the thermal lens (TL) focal length (FL) is found to vary from -1500 mm to -9.78 mm: this corresponds to a linear

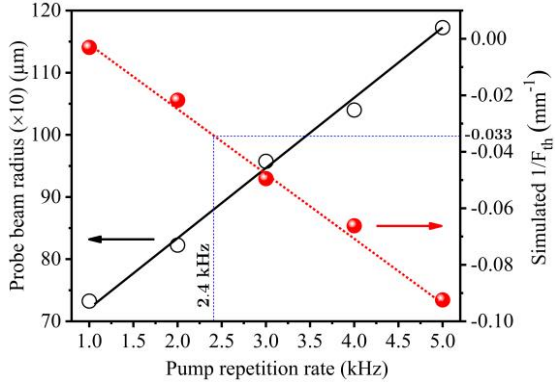

**Fig. 3.** Left: Probe beam radius; right: corresponding calculated thermal lens dioptric power.

increase in dioptric power as shown in Fig. 3.

When thermal lensing only consists of a quadratic phase shift, its main consequence is to modify resonator stability domains. To check if this TL tends to destabilize laser cavity, we simulated our Plano-concave laser cavity using ABCD matrix formalism (Fig. 4). By numerically introducing a diverging TL with FL ranging between -200 mm to -1 mm, the cavity is found to become unstable when the TLFL is smaller than 41 mm in absolute value. This value is comparable but slightly higher than the FL of 24 mm that can be deduced from Fig. 3 at 2.4 kHz where lasing stops.

The presence of a strong TL means that there is room for improvement by resonator design. To test this hypothesis, we added a small 3-mm-FL aspheric lens inside the cavity, located roughly 1-mm away from the cuvette. With this additional lens, ABCD ray matrix calculations show that the cavity becomes stable for a wider range of TL down to -3.2 mm (Fig. 4). Experimentally we observed in these conditions lasing at a repetition rate as high as 14 kHz (Fig 5). Fig. 5 (a) shows the Plano-concave cavity design parameters with the location of thermal lens compensator aspheric lens, cavity mirrors and so on. Fig. 5 (b) shows a comparative plot of laser efficiency *vs*. pump repetition rates with and without correction of TL. The correction with a lens of fixed focal length implies that the resonator is optimized (in terms of matching cavity and pump modes inside the gain medium) for a single repetition rate, which is here around 3 kHz, although lasing is obtained over the whole stability domain up to 14 kHz. A more comprehensive insight can be found in Fig. 5 (c)-(d) with the oscilloscope traces during the first 500 ms of lasing with and without TL compensation. As the repetition rate limitation due to thermal loading is a cumulative effect that builds up pulse after pulse, it is consistent to observe that the first pulses all have the same intensity irrespective of the repetition rates. The intensity however rapidly decreases to a steady-state value where the thermal lens has reached a stable dioptric power. The intensity goes to zero if the thermal lens dioptric power is so strong that it makes the cavity unstable, a situation which however allows the laser to emit a few pulses before shutdown: this situation occurs beyond 2.4 kHz and 14 kHz, respectively without and with TL correction.

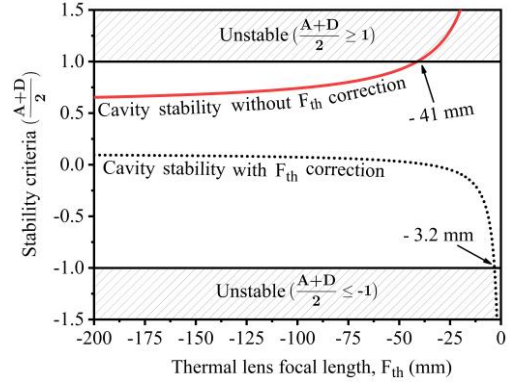

**Fig. 4.** Calculation of the cavity stability criteria c= (A+D)/2 for a Plano-concave laser cavity with a diverging thermal lens; in absence of correction (solid red line), $c > 1$ for thermal lens focal lengths $f_{th} < -41$ mm; with an additional 3-mm positive lens (dotted line), the stability domain is extended to $f_{th} < -3.2$ mm (dotted line).

Interestingly, the TLFL at 14 kHz can be estimated to be -3.3 mm by extending the linear evolution of TL dioptric power *vs*. pump repetition rate (**Fig. S5**), which is very close to the value of -3.2 mm beyond which the cavity was numerically not stable anymore. By extending further the linear relationship between repetition rate and dioptric power (Fig. 3), we can estimate that using a correction lens of shorter FL (for example, – 1.1 mm, repetition rates as high as 40 kHz could theoretically be reached (**Fig. S5**). At such a repetition rate however, triplet issues (together with aberrations of the thermal lens) might start to be important. In order to confirm that the high-repetition rate limit is due to thermal effects, we made the same experiment with a different dye (coumarin-540, from Exciton Luxottica) in the same solvent to keep the thermo-ns pulses) and good pulse-to-pulse stability (RMS <5%) due to an effective photoproduct removal by self-diffusion between pulses. The number of pulses obtainable from a sealed cuvette relates to its total optic properties equal, at a concentration of 1.6 × $10^{-4}$ so that the gain solution had a similar absorption of ~ 85% at 445 nm. The laser efficiency curve using Coumarin is shown in supplementary information (**Fig. S1**). We observed in this case lasing up to 18 kHz with a 3 mm converging lens and a 3 kHz limit without thermal lens correction (**Fig. S6 (a, b)**), very close to the experimental observations obtained with DCM. Unless triplet lifetimes, which are not precisely known in this specific configuration for both materials, would happen to be coincidentally identical, these results suggest the universal nature of the limitation and its thermal origin.

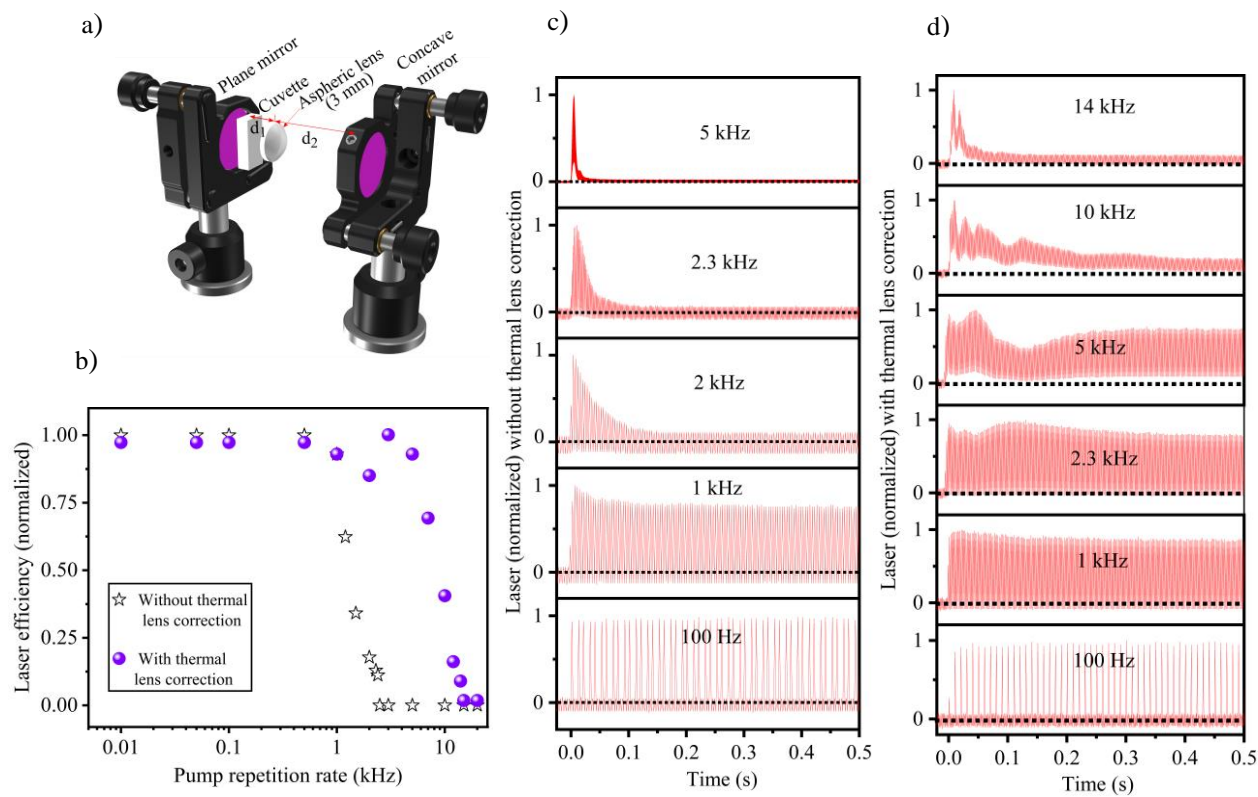

Fig. 5. (a) Plano-concave laser cavity indicating the location of thermal lens compensator aspheric lens ($d_1$=1 mm from cuvette) and output coupler ($d_2$=6.5 mm from aspheric lens). (b) Laser efficiency (normalized) *vs.* pump (30-ns, 83 kW/cm$^2$) repetition rates without and with intracavity thermal lens compensation. (c) Oscilloscope trace of lasing behavior for the first 500 ms at different pump (30-ns, 83 kW/cm$^2$) repetition rates without thermal lens correction and (d) with thermal lens correction.

In conclusion, we have studied the photo stability performance of a circulation-free diode-pumped liquid dye laser. We have shown an extremely high laser longevity (up to 1.2×$10^9$ pulses at 1-kHz for 50 volume, whereas solid-state dye lasers' durability, which is only in the order of thousands of pulses in the same conditions, is instead limited by the pumped volume. The presence of a cumulative effect with a millisecond timescale was evidenced by a rapid drop in laser power observed for repetition rates above the kHz level. This limit is not due to triplet generation but instead to the buildup of a negative thermal lens within the dye cuvette. Correcting the defocus (the non-aberrant part of the TL) can extend the stability domain of the cavity. It was then possible to increase the repetition rate limit up to 14 kHz by simply adding an intracavity lens of fixed focal length. This demonstration opens up opportunities for further improvements in order to reach triplet-limited repetition rates, such as a thermal-lens-insensitive concentric cavity or the insertion of an intracavity variable focal length lens.

## AUTHOR DECLARATION

### Conflict of Interest

The authors have no conflict to disclose.

## DATA AVAILABILITY

The data that support the findings of this study are available from the corresponding author upon reasonable request.

## References

[1] M. Behringer and H. König, PhotonicsViews **17**, 60 (2020).

[2] C.G. Durfee, T. Storz, J. Garlick, S. Hill, J.A. Squier, M. Kirchner, G. Taft, K. Shea, H. Kapteyn, M. Murnane, and S. Backus, CLEO Sci. Innov. CLEO_SI 2012 **20**, 1223 (2012).

[3] A. Hamja, S. Chénais, and S. Forget, J. Appl. Phys. **128**, (2020).

[4] D. Stefańska, M. Suski, A. Zygmunt, J. Stachera, and B. Furmann, Opt. Laser Technol. **120**, 105673 (2019).

[5] O. Burdukova, V. Petukhov, and M. Semenov, Appl. Phys. B Lasers Opt. **124**, 1 (2018).

[6] D. Stefanska, M. Suski, and B. Furmann, Laser Phys. Lett. **14**, (2017).

[7] O. Burdukova, M. Gorbunkov, V. Petukhov, and M. Semenov, Appl. Phys. B Lasers Opt. **123**, 0 (2017).

[8] O.A. Burdukova, M. V. Gorbunkov, V.A. Petukhov, and M.A. Semenov, Laser Phys. Lett. **13**, (2016).

[9] J.H. Gurian, H. Maeda, and T.F. Gallagher, Rev. Sci. Instrum. **81**, 19 (2010).

[10] S.J. Lee, M.J. Choi, Z. Zheng, W.S. Chung, Y.K. Kim, and S. Bin Cho, J. Cosmet. Laser Ther. **15**, 150 (2013).

[11] F. Lahoz, I.R. Martín, J. Gil-Rostra, M. Oliva-Ramirez, F. Yubero, and A.R. Gonzalez-Elipe, Opt. Express **24**, 14383 (2016).

[12] M. Karl, G.L. Whitworth, M. Schubert, C.P. Dietrich, I.D.W. Samuel, G.A. Turnbull, and M.C. Gather, Appl. Phys. Lett. **108**, (2016).

[13] W. Song, A.E. Vasdekis, Z. Li, and D. Psaltis, Appl. Phys. Lett. **94**, 1 (2009).

[14] V.T.N. Mai, A. Shukla, A.M.C. Senevirathne, I. Allison, H. Lim, R.J. Lepage, S.K.M. McGregor, M. Wood, T. Matsushima, E.G. Moore, E.H. Krenske, A.S.D. Sandanayaka, C. Adachi, E.B. Namdas, and S.C. Lo, Adv. Opt. Mater. **8**, (2020).

[15] A.S.D. Sandanayaka, T. Matsushima, F. Bencheikh, K. Yoshida, M. Inoue, T. Fujihara, K. Goushi, J.C. Ribierre, and C. Adachi, Sci. Adv. **3**, 1 (2017).

[16] A.S.D. Sandanayaka, L. Zhao, D. Pitrat, J.C. Mulatier, T. Matsushima, C. Andraud, J.H. Kim, J.C. Ribierre, and C. Adachi, Appl. Phys. Lett. **108**, (2016).

[17] T. Matsushima, S. Yoshida, K. Inada, Y. Esaki, T. Fukunaga, H. Mieno, N. Nakamura, F. Bencheikh, M.R. Leyden, R. Komatsu, C. Qin, A.S.D. Sandanayaka, and C. Adachi, Adv. Funct. Mater. **29**, 1 (2019).

[18] H. Sakata and H. Takeuchi, Appl. Phys. Lett. **92**, (2008).

[19] C. Foucher, B. Guilhabert, A.L. Kanibolotsky, P.J. Skabara, N. Laurand, and M.D. Dawson, Opt. Mater. Express **3**, 584 (2013).

[20] M. Rodriguez, A. Costela, I. Garcia-Moreno, F. Florido, J.M. Figuera, and R. Sastre, Meas. Sci. Technol. **6**, 971 (1995).

[21] V.M. Katarkevich, A.N. Rubinov, T.S. Efendiev, S.S. Anufrik, and M.F. Koldunov, Appl. Opt. **54**, 7962 (2015).

[22] A. Costela, I. García Moreno, C. Gómez, O. García, R. Sastre, A. Roig, and E. Molins, J. Phys. Chem. B **109**, 4475 (2005).

[23] K.M. Abedin, M. Álvarez, A. Costela, I. García-Moreno, O. García, R. Sastre, D.W. Coutts, and C.E. Webb, Opt. Commun. **218**, 359 (2003).

[24] A. Costela, I. García-Moreno, R. Sastre, D.W. Coutts, and C.E. Webb, Appl. Phys. Lett. **79**, 452 (2001).

[25] S. Forget and S. Chénais, *Organic Solid-State Lasers* (Springer Berlin Heidelberg, Berlin, Heidelberg, 2013).

[26] A.S.D. Sandanayaka, T. Matsushima, F. Bencheikh, K. Yoshida, M. Inoue, T. Fujihara, K. Goushi, J.C. Ribierre, and C. Adachi, Sci. Adv. **3**, 1 (2017).

[27] Y. Oyama, M. Mamada, A. Shukla, E.G. Moore, S.C. Lo, E.B. Namdas, and C. Adachi, ACS Mater. Lett. **2**, 161 (2020).

[28] E. McKenna, J. Xue, R. Fan, L. Bintz, R. Dinu, and A. Mickelson, Appl. Opt. **44**, 3063 (2005).

[29] M. Meyer, J.C. Mialocq, and B. Perly, J. Phys. Chem. **94**, 98 (1990).

[30] K. Yagi, S. Shibata, T. Yano, A. Yasumori, M. Yamane, and B. Dunn, J. Sol-Gel Sci. Technol. **4**, 67 (1995).

[31] Z. Zhao, O. Mhibik, M. Nafa, S. Chénais, and S. Forget, Appl. Phys. Lett. **106**, (2015).

[32] H. Rabbani-Haghighi, S. Forget, S. Chénais, and A. Siove, Opt. Lett. **35**, 1968 (2010).

[33] W. Hu, H. Ye, C. Li, Z. Jiang, and F. Zhou, Appl. Opt. **36**, 579 (1997).

[34] S. Chénais, F. Druon, S. Forget, F. Balembois, and P. Georges, Prog. Quantum Electron. **30**, 89 (2006).

[35] W. Xie, S.C. Tam, Y.L. Lam, and Y. Kwon, Opt. Commun. **189**, 337 (2001).

[36] H. Su, Y. Zhang, Y. Zhao, K. Ma, and J. Wang, Opt. Fiber Technol. **51**, 1 (2019).